\documentclass[aps, reprint,showpacs,showkeys,pra,superscriptaddress,longbibliography]{revtex4-1}
\usepackage{dcolumn}
\usepackage[utf8]{inputenc}
\usepackage{amsmath}
\usepackage{amsfonts}
\usepackage{amssymb}
\usepackage{graphicx}
\usepackage{hyperref}
\usepackage{xcolor}

\newcommand{\ket}[1]{\ensuremath{\left|{#1}\right\rangle}}
\newcommand{\bra}[1]{\ensuremath{\left\langle{#1}\right |}}

\begin{document}

\title{Association between quantum paradoxes based on weak values and a realistic interpretation of quantum measurements}

\author{Alice M. Aredes} 
\affiliation{Departamento de F\'isica, Universidade Federal de Minas Gerais, Belo Horizonte, MG 31270-901, Brazil} \affiliation{Instituto de F\'{\i}sica, Universidade Federal do Rio de Janeiro, Caixa Postal 68528, Rio de Janeiro, RJ 21941-972, Brazil}

\author{Pablo L. Saldanha}\email{saldanha@fisica.ufmg.br}\affiliation{Departamento de F\'isica, Universidade Federal de Minas Gerais, Belo Horizonte, MG 31270-901, Brazil}

\date{\today}

\begin{abstract}
Many quantum paradoxes based on a realistic view of weak values were discussed in the last decades. They lead to astonishing conclusions such as the measurement of a spin component of a spin-1/2 particle resulting in $100\hbar$, the separation of a photon from its polarization, and the possibility of having 3 particles in 2 boxes without any 2 particles being in the same box, among others. Here we show that the realistic view of the weak values present in these (and other) works is equivalent to a realistic (and highly controversial) view of quantum measurements, where a measurement reveals the underlying reality of the measured quantity. We discuss that all quantum paradoxes based on weak values simply disappear if we deny these realistic views of quantum measurements and weak values. Our work thus aims to demonstrate the strong assumptions and the corresponding problems present in the interpretation of these quantum paradoxes.
\end{abstract}


\maketitle

\section{Introduction}

The concept of weak value was introduced by Aharonov, Albert, and Vaidman in a seminal paper in 1988 \cite{aharonov88}.  This concept has found many important applications. In the field of quantum metrology, it is associated to the weak value amplification, a technique that can amplify the perturbation one wants to measure, making it detectable by existing detector schemes \cite{hosten08,dixon09,alves15,kim22}. The measurement of weak values is also useful for directly obtaining complex quantities of quantum systems, such as wave functions \cite{lundeen11,mirhosseini14,pan19} and geometrical phases \cite{sjoqvist06,cho19}, among other applications \cite{dressel14,huang19,wagner21}.

Besides the many applications using the weak value concept, many quantum paradoxes based on it were discussed in the past decades. The title of the original paper states one of these paradoxes: ``How the result of a measurement of a component of the spin of a spin-1/2 particle can turn out to be 100$\hbar$'' \cite{aharonov88}. In the 3-box paradox \cite{aharonov91,resch04}, the authors say that there may be a situation with one quantum particle and 3 boxes where, ``in spite of the fact that we have only one particle in the above situation, we find this particle with probability one in any one of the first 2 boxes'' \cite{aharonov91}. In the past-of-a-quantum-particle paradox \cite{vaidman13,danan13}, the authors state that ``the photons tell us that they have been in the parts of the interferometer through which they could not pass'' \cite{danan13}. In the quantum Cheshire cat paradox \cite{aharonov13,denkmayr14}, the authors state that ``in the curious way of quantum mechanics, photon polarization may exist where there is no photon at all'' \cite{aharonov13}. In a sequence of the quantum Cheshire cat paradox \cite{das20,liu20}, it was stated that it would be possible to ``decouple two photons from their respective polarizations and then interchange them during recombination'' \cite{das20}. In the quantum violation of the pigeonhole principle paradox \cite{aharonov16,waegell17,chen19,reznik20}, the authors say that they ``find instances when three quantum particles are put in two boxes, yet no two particles are in the same box'' \cite{aharonov16}.

Something that is not always clearly stated in many of the papers dealing with quantum paradoxes based on weak values is that their paradoxical conclusions depend on a realistic interpretation of the weak values. In other (non-realistic) views, the experimental predictions and experimental results can be understood as simple quantum interference effects. Simple interferometric descriptions, free from paradoxical conclusions, were presented for the paradoxes of the spin measurement resulting in $100\hbar$ \cite{duck89}, of the past of a quantum particle \cite{saldanha14,bartkiewicz15,englert17}, of the quantum Cheshire cat \cite{correa15,atherton15} and its sequence \cite{correa21}, and of the quantum violation of the pigeonhole principle \cite{correa21}, while the 3-box paradox was experimentally implemented with classical light \cite{resch04}, thus also having an interferometric explanation. 

The main objective of the present paper is to present a single argument to criticize all quantum paradoxes cited above, among others (including one proposed here), associating a realistic view of the weak values to the following realistic  interpretation of a quantum measurement \cite{saldanha20}: \textit{A measurement performed on a quantum system reveals the underlying ontological value of the measured quantity, that continues the same after the measurement is performed}. This interpretation is highly controversial, generating numerous paradoxes. For instance, if a measurement reveals a pre-existing value for the measured quantity, this value must depend on which other compatible observers are simultaneously measured, due to quantum contextuality \cite{budroni22}.  In the present work, we conclude that all the cited paradoxes disappear if we simply deny this realistic and controversial interpretation of a quantum measurement, also denying a realistic view of the weak values. In this sense, the cited paradoxes can be seen as demonstrations that realistic interpretations of quantum measurements and of weak values lead to inconsistencies, not as demonstrations of astonishing behaviors of Nature.

\section{General Argument}

In a weak measurement procedure, a quantum system is pre-selected in a state $\ket{\psi_i}$ and post-selected in a state $\ket{\psi_f}$. Between the pre- and post-selection, the system interacts with a probe (which is also a quantum system) that extracts some information. The weakness of the interaction means that the quantum system induces a small change on the probe state. So, considering the probe system as a pointer, the central position of the pointer wave function is displaced by an amount much smaller than its initial quantum uncertainty. This displacement is proportional to the real part of the weak value $\langle O\rangle_w$ of the observable $O$ from the system that rules the system-pointer interaction, defined as \cite{aharonov88,dressel14}
\begin{equation}\label{weak}
	\langle O\rangle_w=\frac{\bra{\psi_f}O\ket{\psi_i}}{\langle\psi_f|\psi_i\rangle}.
\end{equation}
The weak value $\langle O\rangle_w$ may assume values outside the eigenvalue spectrum of the observable $O$, including complex values. Since the real part of $\langle O\rangle_w$ is associated to the shift of the measuring device pointer, $\langle O\rangle_w$ is sometimes considered to be a quantity associated to the result of a measurement. However, since in a weak measurement procedure the shift of the pointer must be much smaller than its quantum uncertainty, many repetitions of the protocol are necessary to experimentally extract the weak value \cite{aharonov88,dressel14}. In fact, for an accurate determination of the weak value, infinite repetitions are necessary \cite{haapasalo11}.

Our method for associating a realistic interpretation of the weak values to the cited realistic interpretation of quantum measurements applies to observables $O$ that can be written as the sum of an observable $P$ from which the pre-selected state is an eigenvector with eigenvalue $p$ and an observable $Q$ from which the post-selected state is an eigenvector with eigenvalue $q$:
\begin{equation}\label{cond}
	O=P+Q,\;\mathrm{with}\;\;\;P\ket{\psi_i}=p\ket{\psi_i},\;\;Q\ket{\psi_f}=q\ket{\psi_f}.
\end{equation}
For observables $O$ that obey Eq. (\ref{cond}), the weak value can be written as
\begin{equation}
	\langle O\rangle_w=\langle P\rangle_w+\langle Q\rangle_w=\frac{\bra{\psi_f}P\ket{\psi_i}}{\langle\psi_f|\psi_i\rangle}+\frac{\bra{\psi_f}Q\ket{\psi_i}}{\langle\psi_f|\psi_i\rangle}=p+q,
\end{equation}
where Eq. (\ref{cond}) and the fact that $q$ is real were used. In our argument, the operators $P$ and $Q$ may commute or not. For most of the observables $O$ relevant in the quantum paradoxes to be treated in the next section, their decomposition as in Eq. (\ref{cond}) are done with operators $P$ and $Q$ that do not commute.

In this work we use tildes to represent physical quantities with an object reality, under the realistic assumptions we discuss. For instance, if $S_z$ is the operator representing the $z$ component of the spin of a spin-1/2 particle, by writing $\tilde{S}_z=\hbar/2$ we are assuming that the $z$ component of the particle spin has the ontological value $\hbar/2$ at that time. The realistic assumption that leads to all the quantum paradoxes cited before \cite{aharonov88,aharonov91,resch04,vaidman13,danan13,aharonov13,denkmayr14,das20,liu20,aharonov16,waegell17,chen19,reznik20} is that the weak value $\langle O\rangle_w$ of an operator $O$ reveals the objective reality of the physical quantity $\tilde{O}$ associated to this operator at a time between the pre- and post-selections. Under this realistic view of the weak values, for observables that can be written as in Eq. (\ref{cond}) we have $\tilde{P}=\langle P\rangle_w=p$, $\tilde{Q}=\langle Q\rangle_w=q$, and $\tilde{O}=\langle O\rangle_w=p+q$. 

Let us now consider the cited realistic interpretation of quantum measurements in this situation. Since it is assumed that a measurement reveals the underlying ontological value of the measured quantity, considering that the post-selection of the state $\ket{\psi_f}$ includes a measurement of the observable $Q$ resulting in the eigenvalue $q$, we conclude that we have an objective value $\tilde{Q}=q$ for this physical quantity before this measurement. With the assumption that the objective value of the measured quantity continues the same after the measurement is performed, considering that the pre-selection of the state $\ket{\psi_i}$ includes a measurement of the observable $P$ resulting in the eigenvalue $p$, we conclude that $\tilde{P}=p$ after this measurement. So, at a time between the pre- and post-selection, we have $\tilde{P}=p$, $\tilde{Q}=q$, and, consequently, $\tilde{O}=\tilde{P}+\tilde{Q}=p+q$. Note that the equality $\tilde{O}=\tilde{P}+\tilde{Q}$ only holds because both physical quantities $\tilde{P}$ and $\tilde{Q}$ have definite values at the same time in this example under our assumptions. We thus see that, when the observable $O$ can be written as in Eq. (\ref{cond}), the attribution of a physical reality to the weak value $\langle O\rangle_w$ is equivalent to adopting the cited realistic interpretation of quantum measurements, since both assumptions lead to the same objective value for the physical quantity $\tilde{O}$: $\tilde{O}=p+q$. 

In the following, we show that all observables used in the cited quantum paradoxes based on weak values \cite{aharonov88,aharonov91,resch04,vaidman13,danan13,aharonov13,denkmayr14,das20,liu20,aharonov16,waegell17,chen19,reznik20} can be written as in Eq. (\ref{cond}). Since the cited realistic interpretation of quantum measurements is highly controversial, a reasonable way to avoid all the cited quantum paradoxes is to deny this realistic interpretation of quantum measurements, also denying the realistic view of the weak values.

\section{Quantum Paradoxes Revisited}

\subsection{Measurement of spin resulting in 100$\hbar$}

Let us start discussing the paradox of the original paper from Aharonov, Albert, and Vaidman, dealing with a spin-1/2 particle \cite{aharonov88}. The particle is pre-selected by a Stern-Gerlach apparatus with the magnetic field in a direction $\xi$ in the $xz$ plane that makes an angle $\alpha$ with the $x$ direction, obtaining a spin component $+\hbar/2$ in this direction $\xi$. The post-selection is made by a Stern-Gerlach apparatus with magnetic field in the $x$ direction, obtaining a spin component $+\hbar/2$ in the direction $x$. The weak value of the $z$ component of spin can be computed by Eq. (\ref{weak}) and the result is $\langle S_{z}\rangle_w=(\hbar/2)\tan(\alpha/2)$ \cite{aharonov88}. For $\alpha=179.43^\circ$, we have $\langle S_{z}\rangle_w\approx100\hbar$. If, between the pre- and post-selection, the quantum particle interacts with a small nonuniform magnetic field in the $z$ direction, its average momentum gain will be the same as the one of a particle with spin $100\hbar$ (and the same gyromagnetic ratio) \cite{aharonov88}. This fact may lead to the attribution of a physical reality to the $z$ component of the particle spin as being $\tilde{S}_z=\langle S_{z}\rangle_w\approx100\hbar$. In this realistic view, the measurement of the particle momentum deviation would be a measurement of the particle spin component in the $z$ direction between the two Stern-Geralch apparatuses. However, this \textit{weak measurement} procedure only works if the momentum deviation is much smaller than the initial momentum quantum uncertainty of the particle in the $z$ direction, such that the result can be completely understood in interferometric terms \cite{duck89}. The post-selection selects a portion of the initial momentum wave function with higher values of momentum, such that the average momentum of the selected wave function has a relatively large value, but only contains momentum components that were already present in the initial distribution. So, there is no need to attribute a physical reality to the weak value of the $z$ component of the particle spin \cite{duck89}.  

Let us now see what are the conclusions about the value of the $z$ component of the particle spin between the Stern Gerlach measurements using the cited realistic interpretation of quantum measurements. Under the pre- and post-selection, we attribute objective values to the spin components in the directions of the apparatuses magnetic fields at a time between the measurements: $\tilde{S}_\xi=\hbar/2$, $\tilde{S}_x=\hbar/2$. As depicted in Fig. 1, these values for these spin components imply in a  $z$ component of the spin $\tilde{S}_z=(\hbar/2)\tan(\alpha/2)$, equal to $\langle S_{z}\rangle_w$. The equivalence between $\tilde{S}_z$ and $\langle S_{z}\rangle_w$ occurs because we have $S_z=[S_\xi-\cos(\alpha)S_x]/\sin(\alpha)$, which is in the form of Eq. (\ref{cond}) with $O=S_z$, $P=S_\xi/\sin(\alpha)$, and $Q=-S_x/\tan(\alpha)$. The realistic interpretation of the particle spin as the vectors represented in Fig. 1 is certainly controversial, and leads to the same conclusions as the realistic view of the weak values, with the $z$ component of spin having the value $\tilde{S}_z=(\hbar/2)\tan(\alpha/2)$ at a time between the measurements. Since both realistic interpretations lead to the same conclusions here and in the other cases we treat, we say they are equivalent. 

\begin{figure}
  \centering
    \includegraphics[width=8.5cm]{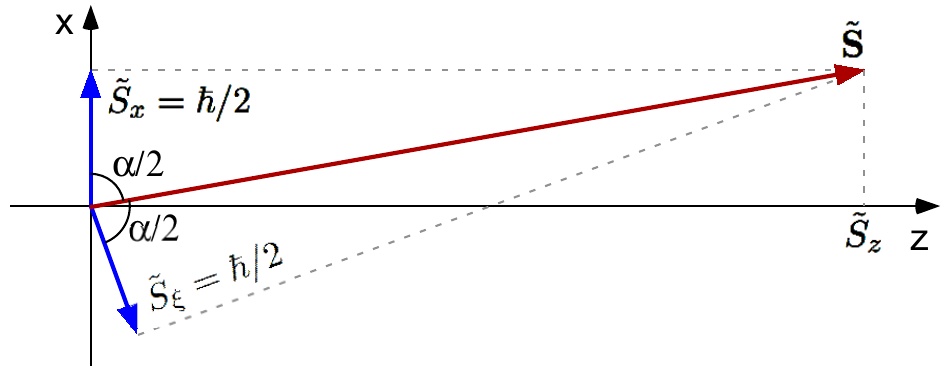}
  \caption{Realistic view of the spin components in the paradox of a spin measurement resulting in $100\hbar$ \cite{aharonov88}.}
\end{figure}


\subsection{3-box paradox}

Let us now consider the 3-box paradox, where a realistic view of the weak values leads to the conclusion that a quantum particle can be found at 2 different boxes with probability 1 in each box \cite{aharonov91}. A quantum particle is prepared in a superposition state of being in three orthogonal states $\ket{A}$, $\ket{B}$, and $\ket{C}$, which can be considered as the states of the particle inside boxes labeled $A$, $B$, and $C$ respectively. The pre-selected state is $\ket{\psi_i}=(\ket{A}+\ket{B}+\ket{C})/\sqrt{3}$ and the post-selected state is $\ket{\psi_f}=(\ket{A}+\ket{B}-\ket{C})/\sqrt{3}$. Projectors $\Pi_j=\ket{j}\bra{j}$ (with $j=\{A,B,C\}$) are associated to the presence of the particle in box $j$, with an eigenvalue 1 representing the presence and an eigenvalue 0 the absence of the particle in the corresponding box. For this configuration, we compute the following weak values using Eq. (\ref{weak}): $\langle \Pi_{A}\rangle_w=1$, $\langle \Pi_{B}\rangle_w=1$, $\langle \Pi_{C}\rangle_w=-1$. By adopting a realistic view of the weak values, the authors conclude that the probability of the particle to be found in box $A$ is 1 (since $\langle \Pi_{A}\rangle_w=1$), as well as the probability that the particle to be found in box $B$ (since $\langle \Pi_{B}\rangle_w=1$) \cite{aharonov91}. 

There are many different ways to decompose the operators $\Pi_A$, $\Pi_B$, and $\Pi_C$ of the 3-box paradox in the form of  Eq. (\ref{cond}). We only show one possible decomposition for each operator below, in the basis $\{\ket{A},\ket{B},\ket{C}\}$ in matrix form. For $\Pi_A$, with $O=\Pi_A$, we can write
\begin{equation}\nonumber
	P=\begin{pmatrix}
5/3 & -7/6 & 0\\
-7/6 & 7/6 & 1/2\\
0 & 1/2 & 0
\end{pmatrix},\;
Q=\begin{pmatrix}
-2/3 & 7/6 & 0\\
7/6 & -7/6 & -1/2\\
0 & -1/2 & 0
\end{pmatrix},
\end{equation}
$p=q=1/2$. The physical interpretation of the above observables is not as simple as in the previous case that considered spin components, but the idea is the same. By assuming the cited realistic interpretation of quantum measurements, we obtain an ontological value for the presence of the particle in box $A$ given by $\tilde{\Pi}_A=\tilde{P}+\tilde{Q}=p+q=1$, in the same way as by assuming a realistic view of the weak value $\langle \Pi_{A}\rangle_w=1$. For $\Pi_B$, with $O=\Pi_B$ in Eq. (\ref{cond}), we can write
\begin{equation}\nonumber
	P=\begin{pmatrix}
-1/2 & 1 & 1/2\\
1 & 0 & 0\\
1/2 & 0 & 1/2
\end{pmatrix},\;
Q=\begin{pmatrix}
1/2 & -1 & -1/2\\
-1 & 1 & 0\\
-1/2 & 0 & -1/2
\end{pmatrix},
\end{equation}
$p=1$, $q=0$. By assuming the cited realistic interpretation of quantum measurements, we obtain  $\tilde{\Pi}_B=p+q=1=\langle \Pi_{B}\rangle_w$. For $\Pi_C$, with $O=\Pi_C$ in Eq. (\ref{cond}), we can write
\begin{equation}\nonumber
	P=\begin{pmatrix}
1/2 & 1/2 & -1/2\\
1/2 & 1/2 & -1/2\\
-1/2 & -1/2 & 3/2
\end{pmatrix},\;
Q=\begin{pmatrix}
-1/2 & -1/2 & 1/2\\
-1/2 & -1/2 & 1/2\\
1/2 & 1/2 & -1/2
\end{pmatrix},
\end{equation}
$p=1/2$, $q=-3/2$. We obtain $\tilde{\Pi}_C=p+q=-1=\langle \Pi_{C}\rangle_w$ under the cited assumptions. We see that the cited realistic interpretation of quantum measurements leads to the same conclusions as the realistic view of the weak values.

As mentioned before, the relevant operators for the paradox can be decomposed as in Eq.  (\ref{cond}) in different ways. Consider that an operator $O$ can be decomposed as in Eq. (\ref{cond}), but also as $O=P'+Q'$, with $P'\ket{\psi_i}=p'\ket{\psi_i}$ and $Q'\ket{\psi_f}=q'\ket{\psi_f}$. Obviously, we must have $p+q=p'+q'$. We can consider that the pre-selection is performed by the measurement of the observable $P$, such that we have $\tilde{P}=p$ under the realistic view of quantum measurements we consider in this work, or by the measurement of the operator  $P'$, such that we have $\tilde{P}'=p'$, or by both measurements, associating ontological values to both $\tilde{P}$ and $\tilde{P}'$. Similar considerations can be done for the post-selection. In any case, we have $\tilde{O}=\langle O \rangle_w$=p+q=p'+q', such that realistic interpretation of quantum measurements leads to the same conclusion as the realistic view of the weak values.

\subsection{Quantum Cheshire cat}

The quantum Cheshire cat paradox considers the interferometer depicted in Fig. 2 \cite{aharonov13}. A photon with horizontal polarization is sent to the interferometer, such that after the beam splitter BS$_1$ the pre-selected state is $\ket{\psi_i}=(i\ket{L}+\ket{R})\ket{H}/\sqrt{2}$, where $\ket{L}$ and $\ket{R}$ represent the possible photon paths in the interferometer, as depicted in the figure, and $\ket{H}$ and $\ket{V}$ represent horizontal and vertical linear polarizations for the photon, respectively. The phase shifter PS includes a phase $\pi/2$ in the corresponding path and the half-wave plate HWP rotates the polarization from $\ket{H}$ to $\ket{V}$, such that a photon detection by detector D$_1$ corresponds to the post-selection of the state $\ket{\psi_f}=(\ket{L}\ket{H}+\ket{R}\ket{V})/\sqrt{2}$ inside the interferometer. The presence of the photon in each path of the interferometer is associated to the observables $\Pi_L=\ket{L}\bra{L}$ and $\Pi_R=\ket{R}\bra{R}$. The polarization in the circular basis $\ket{\pm}=(\ket{H}\pm i\ket{V})$ is associated to the observable $\sigma_z=\ket{+}\bra{+}-\ket{-}\bra{-}$. The circular polarization in each path is associated to the observables $\sigma_z^{(L)}=\Pi_L \otimes \sigma_z$ and $\sigma_z^{(R)}=\Pi_R \otimes \sigma_z$. The following weak values are found with the use of  Eq. (\ref{weak}): $\langle \Pi_{L}\rangle_w=1$, $\langle \Pi_{R}\rangle_w=0$, $\langle \sigma_z^{(L)}\rangle_w=0$, $\langle \sigma_z^{(R)}\rangle_w=1$. By adopting a realistic view of the weak values, the authors conclude that the photon propagates through path $L$ (since $\langle \Pi_{L}\rangle_w=1$ and $\langle \Pi_{R}\rangle_w=0$), but its polarization propagates through path $R$ (since $\langle \sigma_z^{(L)}\rangle_w=0$ and $\langle \sigma_z^{(R)}\rangle_w=1$) \cite{aharonov13}. But we reinforce that the theoretical predictions \cite{aharonov13} and experimental results \cite{denkmayr14} of the quantum Cheshire cat effect can be explained as simple quantum interference effects \cite{correa15,atherton15}. 

\begin{figure}
  \centering
    \includegraphics[width=8.5cm]{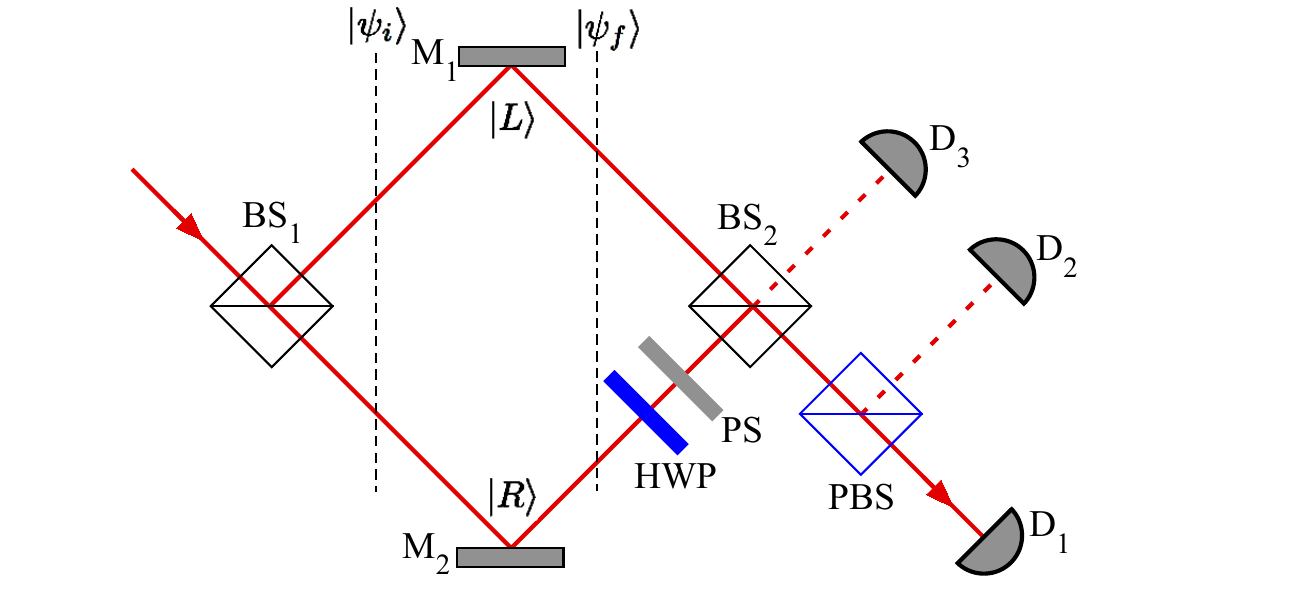}
  \caption{Scheme of the quantum Cheshire cat paradox \cite{aharonov13}. BS$_1$ and BS$_2$ are beam splitters, M$_1$ and M$_2$ are mirrors, PBS is a polarizing beam splitter, PS is a phase shifter, HWP is a half-wave plate, D$_1$, D$_2$, and D$_3$ are photon detectors.}
\end{figure}

To show that the cited realistic interpretation of quantum measurements lead to the same conclusions as the realistic view of the weak values in the quantum Cheshire cat paradox, such that these realistic assumptions may be considered equivalent, we only need to show that each of the observables $\Pi_L$, $\Pi_R$, $\sigma_z^{(L)}$, and $\sigma_z^{(R)}$ can be decomposed as in Eq. (\ref{cond}). We do this in Appendix \ref{ap-cheshire}.

\subsection{Past of a quantum particle}

The paradox regarding the past of a quantum particle uses a scheme like the one depicted in Fig. 3, with a nested Mach-Zehnder interferometer \cite{vaidman13,danan13}. A photon is sent to the interferometer with beam splitters projected such that its pre-selected state inside the interferometer is $\psi_i=(\ket{A}+i\ket{B}+\ket{C})/\sqrt{3}$. A state $\ket{j}$ represents the photon in a path that includes mirror $j$. A photon detection by the detector D post-select the quantum state $\psi_f=(\ket{A}-i\ket{B}+\ket{C})/\sqrt{3}$. This configuration implies that there is destructive interference in the inner interferometer for light exiting in the direction of mirror F. The weak values of the projectors $\Pi_A=\ket{A}\bra{A}$, $\Pi_B=\ket{B}\bra{B}$, and $\Pi_E=\ket{E}\bra{E}$ are found to be $\langle \Pi_A\rangle_w=1$, $\langle \Pi_B\rangle_w=-1$, $\langle \Pi_E\rangle_w=0$. By using a realistic view of the weak values, the authors conclude that the photon passes through path A  (since $\langle \Pi_A\rangle_w=1$), but not through path E (since $\langle \Pi_E\rangle_w=0$), which is impossible. It is important to stress that both the theoretical prediction \cite{vaidman13} and the experimental implementation \cite{danan13} of this paradox can be described as simple interference effects \cite{saldanha14,bartkiewicz15,englert17}.  

\begin{figure}
  \centering
    \includegraphics[width=8.5cm]{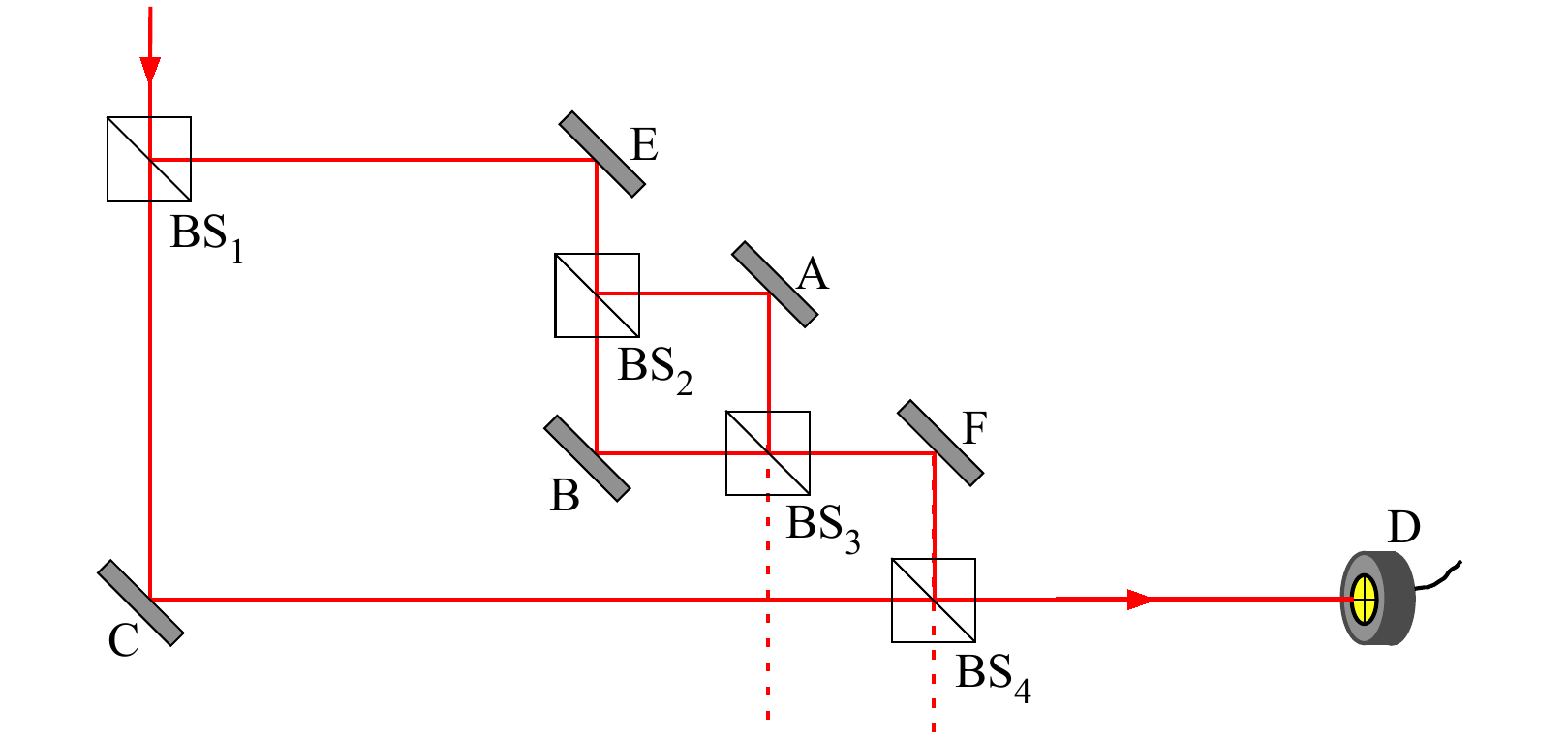}
  \caption{Scheme of the paradox involving the past of a quantum particle \cite{danan13}. BS$_1$, BS$_2$, BS$_3$, and BS$_4$ are beam splitters. A, B, C, E, and F are mirrors. D is a photon detector.}
\end{figure}

In Appendix \ref{ap-past} we show that each of the observables $\Pi_A$, $\Pi_B$, and $\Pi_E$ of the paradox involving the past of a quantum particle can be decomposed as in Eq. (\ref{cond}). In this way, the cited realistic assumption regarding quantum measurements leads to the same conclusions as the realistic view of the weak values, so that these realistic assumptions are equivalent.

\subsection{Quantum pigeonhole paradox}

In the quantum violation of the pigeonhole principle \cite{aharonov16}, there is a system of 3 particles, each of which can be in orthogonal states $\ket{L}$ or $\ket{R}$,  associated to the presence in one of two boxes $L$ and $R$, as well as in superposition states. The pre-selected state is $\psi_i=\ket{x}_1\ket{x}_2\ket{x}_3$, with $\ket{x}=(\ket{L}+\ket{R})/\sqrt{2}$. The post-selected state is $\psi_f=\ket{+}_1\ket{+}_2\ket{+}_3$, with $\ket{+}=(\ket{L}+i\ket{R})/\sqrt{2}$. The observable $\Pi^\mathrm{same}_{1,2}=\ket{L}_1\ket{L}_2\bra{L}_1\bra{L}_2+\ket{R}_1\ket{R}_2\bra{R}_1\bra{R}_2$ is associated to the presence of the particles 1 and 2 in the same box. The eigenvalue 1 corresponds to the two particles being in the same box, while the eigenvalue 0 corresponds to the two particles being in different boxes. Using Eq. (\ref{weak}), the weak value of this observable is $\langle \Pi^\mathrm{same}_{1,2}\rangle_w=0$. Since the pre- and post-selected states are symmetric under the exchange of any two particles, we also have $\langle \Pi^\mathrm{same}_{1,3}\rangle_w=\langle \Pi^\mathrm{same}_{2,3}\rangle_w=0$, with obvious notation. So, a realistic view of the weak values imply that no two particles are in the same box at times between the pre- and post-selection, even if we have 3 particles and 2 boxes. But it is important to stress that the theoretical predictions \cite{aharonov16} and experimental results \cite{waegell17,chen19} regarding this paradox can be understood as simple interference effects, without such paradoxical conclusions \cite{correa21}.

To show that the cited realistic interpretation of quantum measurements lead to the same conclusions as the realistic view of the weak values in the quantum violation of the pigeonhole principle, so that these realistic assumptions may be considered equivalent, we only need to show that each of the observables $\Pi^\mathrm{same}_{1,2}$, $\Pi^\mathrm{same}_{1,3}$, and $\Pi^\mathrm{same}_{2,3}$ can be decomposed as in Eq. (\ref{cond}). We do this in Appendix \ref{ap-pigeonhole}.

\subsection{A novel paradox with a spin-2 particle}

In this subsection we propose and criticize a novel paradox based on a realistic view of the weak values. The paradox is inspired in the situation used in Ref. \cite{saldanha20} to criticize the realistic interpretation of quantum measurements we are discussing here. Consider that a spin-2 particle is pre-selected in the sate $\ket{2\hbar}_x$, an eigenstate of $S_x$ with eigenvalue $2\hbar$, and post-selected in the sate $\ket{2\hbar}_z$, an eigenstate of $S_z$ with eigenvalue $2\hbar$ ($S_i$ is the $i$ component of the particle spin). It can be readily shown that the weak value of $S_y^2$ is given by $\langle S_y^2\rangle_w=-2\hbar^2$. We could argue that in this situation the square of the $y$ component of the particle spin has a negative value, demonstrating an astonishing behavior of Nature and contradicting the notion that the square of a real quantity must be positive or zero.  But this conclusion would be based on a realistic view of the weak values, that we can simply deny. 

The same paradoxical conclusion is achieved with the cited realistic interpretation of quantum measurements, as discussed in Ref. \cite{saldanha20}. We can write $S_y^2=S^2-S_x^2-S_z^2$, with any state of a spin-2 particle being an eigenstate of $S^2$ with eigenvalue $6\hbar^2$. The operator $S_y^2$ can be decomposed as in Eq. (\ref{cond}), with $O=S_y^2$, $P=S^2-S_x^2$, and $Q=-S_z^2$. So, if we simultaneously attribute physical reality to the quantities $\tilde{S}^2=6\hbar^2$, $\tilde{S}_x^2=4\hbar^2$, and $\tilde{S}_z^2=4\hbar^2$ at a time between the pre- and post-selection, due to the results of the performed measurements, we conclude that $\tilde{S}_y^2=\tilde{S}^2-\tilde{S}_x^2-\tilde{S}_z^2=-2\hbar^2$.

\section{Conclusion}

The assumption that a quantum measurement reveals the underlying reality of the measured quantity is certainly controversial, incapable of describing quantum phenomena without generating numerous paradoxes \cite{saldanha20}. We have shown that this realistic assumption regarding quantum measurements leads to the same conclusions as the realistic view of the weak values in many quantum paradoxes described in the literature \cite{aharonov88,aharonov91,resch04,vaidman13,danan13,aharonov13,denkmayr14,das20,liu20,aharonov16,waegell17,chen19,reznik20} and in one proposed here. So, instead of assuming the validity of the bizarre behaviors of Nature described in these works, one may simply deny the cited realistic interpretation of a quantum measurement, also denying the realistic view of the weak values. Quantum mechanics is a very rich, complex, and difficult subject. We believe that the numerous quantum paradoxes based on a realistic view of the weak values present in the literature tend to obscure, rather than to clarify, the understanding of quantum phenomena. After all, we can interpret these paradoxes as showing that these realistic views of quantum measurements and weak values are not reasonable. We hope our work contributes to reinforce this point of view.

This work was supported by the Brazilian agencies CNPq (Conselho Nacional de Desenvolvimento Científico e Tecnológico), CAPES (Coordenação de Aperfeiçoamento de Pessoal de Nível Superior), and FAPEMIG (Fundação de Amparo à Pesquisa do Estado de Minas Gerais).

\appendix

\section{Observables of the quantum Cheshire cat paradox} \label{ap-cheshire}

Here we show that each of the observables $\Pi_L$, $\Pi_R$, $\sigma_z^{(L)}$, and $\sigma_z^{(R)}$ of the quantum Cheshire cat paradox can be decomposed as in Eq. (\ref{cond}), in the basis $\{\ket{L}\ket{H},\ket{L}\ket{V},\ket{R}\ket{H},\ket{R}\ket{V}\}$ in matrix form. One possible decomposition for $\Pi_L$, with $O=\Pi_L$ in Eq. (\ref{cond}), is
\begin{equation}\nonumber
	P=\begin{pmatrix}
2 & 1 & -i & -1 \\
1 & 1 & -i & -1 \\
i & i & 2 & -i \\
-1 & -1 & i & 1 
\end{pmatrix},\;
Q=\begin{pmatrix}
-1 & -1 & i & 1 \\
-1 & 0 & i & 1 \\
-i & -i & -2 & i \\
1 & 1 & -i & -1 
\end{pmatrix},
\end{equation}
$p=1$, $q=0$, which results in $\tilde{\Pi}_L=1=\langle\Pi_{L}\rangle_w$ with the cited realistic interpretation of quantum measurements. For $\Pi_R$, with $O=\Pi_R$ in Eq. (\ref{cond}), we can write
\begin{equation}\nonumber
	P=\begin{pmatrix}
0 & 1 & i & 1 \\
1 & 1 & -i & -1 \\
-i & i & 0 & i \\
1 & -1 & -i & 1 
\end{pmatrix},\;
Q=\begin{pmatrix}
0 & -1 & -i & -1 \\
-1 & -1 & i & 1 \\
i & -i & 1 & -i \\
-1 & 1 & i & 0 
\end{pmatrix},
\end{equation}
$p=1$, $q=-1$, which results in $\tilde{\Pi}_R=0=\langle\Pi_{R}\rangle_w$ under the realistic assumptions. For $\sigma_z^{(L)}$, with $O=\sigma_z^{(L)}$ in Eq. (\ref{cond}), we can write
\begin{equation}\nonumber
	P=\begin{pmatrix}
1 & 0 & 0 & 0 \\
0 & 1 & 0 & i \\
0 & 0 & 1 & 0 \\
0 & -i & 0 & 1 
\end{pmatrix},\;
Q=\begin{pmatrix}
-1 & -i & 0 & 0 \\
i & -1 & 0 & -i \\
0 & 0 & -1 & 0 \\
0 & i & 0 & -1 
\end{pmatrix},
\end{equation}
$p=1$, $q=-1$, what results in $\tilde{\sigma}_z^{(L)}=0=\langle\sigma_{z}^{(L)}\rangle_w$ under the realistic assumptions. For $\sigma_z^{(R)}$, with $O=\sigma_z^{(R)}$ in Eq. (\ref{cond}), we can write
\begin{equation}\nonumber
	P=\begin{pmatrix}
1 & 1 & 0 & -1 \\
1 & 1 & -i & -1 \\
0 & i & 1 & -i \\
-1 & -1 & i & 1 
\end{pmatrix},\;
Q=\begin{pmatrix}
-1 & -1 & 0 & 1 \\
-1 & -1 & i & 1 \\
0 & -i & -1 & 0 \\
1 & 1 & 0 & -1 
\end{pmatrix},
\end{equation}
$p=1$, $q=0$, what results in $\tilde{\sigma}_z^{(R)}=1=\langle\sigma_{z}^{(R)}\rangle_w$ under the realistic assumptions.

\section{Observables of the paradox regarding the past of a quantum particle} \label{ap-past}

Here we show that the relevant observables of the paradox regarding the past of a quantum particle ($\Pi_A$, $\Pi_B$, and $\Pi_E$) can be decomposed as in Eq. (\ref{cond}), in the basis $\{\ket{A}, \ket{B}, \ket{C}\}$ in matrix form. One possible decomposition for $\Pi_A$, with $O=\Pi_A$ in Eq. (\ref{cond}), is
\begin{equation}\nonumber
	P=\frac{1}{6}\begin{pmatrix}
2 & 0 & 1 \\
0 & 0 & 3i \\
1 & -3i & -1 
\end{pmatrix},\;
Q=\frac{1}{6}\begin{pmatrix}
4 & 0 & -1 \\
0 & 0 & -3i \\
-1 & 3i & 1 
\end{pmatrix},\;
\end{equation}
$p=q=1/2$, what results in $\tilde{\Pi}_A=1=\langle\Pi_{A}\rangle_w$ with the cited realistic interpretation of quantum measurements. For $\Pi_B$, with with $O=\Pi_B$ in Eq. (\ref{cond}), we can write
\begin{equation}\nonumber
	P=\frac{1}{6}\begin{pmatrix}
1 & 3i & -1 \\
-3i & 3 & -3i \\
-1 & 3i & 1 
\end{pmatrix},\;
Q=\frac{1}{6}\begin{pmatrix}
-1 & -3i & 1 \\
3i & 3 & 3i \\
1 & -3i & -1 
\end{pmatrix},\;
\end{equation}
$p=q=-1/2$, what results in $\tilde{\Pi}_B=-1=\langle\Pi_{B}\rangle_w$ with the realistic interpretation of quantum measurements. $\Pi_E$ is an eigenstate of $\ket{\psi_f}$ with eigevalue $0$, so that it can be readily written in the form of Eq.  (\ref{cond}) with $O=Q=\Pi_E$, $P=0$, $p=q=0$, what results in $\tilde{\Pi}_E=0=\langle\Pi_{E}\rangle_w$ under the realistic assumptions.

\section{Observables of the quantum pigeonhole paradox} \label{ap-pigeonhole}

Here we show that each of the observables $\Pi^\mathrm{same}_{i,j}$ of the quantum pigeonhole paradox can be decomposed as in Eq. (\ref{cond}). All these operators have the same form in the subspace of particles $i$ and $j$ in the basis $\{\ket{L}_i\ket{L}_j,\ket{L}_i\ket{R}_j,\ket{R}_i\ket{L}_j,\ket{R}_i\ket{R}_j\}$ in matrix form. One  possible decomposition, with $O=\Pi^\mathrm{same}_{i,j}$ in Eq. (\ref{cond}), is
\begin{eqnarray}\nonumber
	P&=&\frac{1}{4}\begin{pmatrix}
5 & 5 & -4-i & -2+i \\
5 & 5 & -6+5i & -5i \\
-4+i & -6-5i & 13 & 1+4i \\
-2-i & 5i & 1-4i & 5 
\end{pmatrix},\;\\\nonumber
Q&=&\frac{1}{4}\begin{pmatrix}
-1 & -5 & 4+i & 2-i \\
-5 & -5 & 6-5i & 5i \\
4-i & 6+5i & -13 & -1-4i \\
2+i & -5i & -1+4i & -1 
\end{pmatrix},\;
\end{eqnarray}
$p=1$, $q=-1$, which results in $\tilde{\Pi}^\mathrm{same}_{i,j}=0=\langle\Pi^\mathrm{same}_{i,j}\rangle_w$ with the cited realistic interpretation of quantum measurements.


\begin{thebibliography}{35}%
\makeatletter
\providecommand \@ifxundefined [1]{%
 \@ifx{#1\undefined}
}%
\providecommand \@ifnum [1]{%
 \ifnum #1\expandafter \@firstoftwo
 \else \expandafter \@secondoftwo
 \fi
}%
\providecommand \@ifx [1]{%
 \ifx #1\expandafter \@firstoftwo
 \else \expandafter \@secondoftwo
 \fi
}%
\providecommand \natexlab [1]{#1}%
\providecommand \enquote  [1]{#1}%
\providecommand \bibnamefont  [1]{#1}%
\providecommand \bibfnamefont [1]{#1}%
\providecommand \citenamefont [1]{#1}%
\providecommand \href@noop [0]{\@secondoftwo}%
\providecommand \href [0]{\begingroup \@sanitize@url \@href}%
\providecommand \@href[1]{\@@startlink{#1}\@@href}%
\providecommand \@@href[1]{\endgroup#1\@@endlink}%
\providecommand \@sanitize@url [0]{\catcode `\\12\catcode `\$12\catcode
  `\&12\catcode `\#12\catcode `\^12\catcode `\_12\catcode `\%12\relax}%
\providecommand \@@startlink[1]{}%
\providecommand \@@endlink[0]{}%
\providecommand \url  [0]{\begingroup\@sanitize@url \@url }%
\providecommand \@url [1]{\endgroup\@href {#1}{\urlprefix }}%
\providecommand \urlprefix  [0]{URL }%
\providecommand \Eprint [0]{\href }%
\providecommand \doibase [0]{http://dx.doi.org/}%
\providecommand \selectlanguage [0]{\@gobble}%
\providecommand \bibinfo  [0]{\@secondoftwo}%
\providecommand \bibfield  [0]{\@secondoftwo}%
\providecommand \translation [1]{[#1]}%
\providecommand \BibitemOpen [0]{}%
\providecommand \bibitemStop [0]{}%
\providecommand \bibitemNoStop [0]{.\EOS\space}%
\providecommand \EOS [0]{\spacefactor3000\relax}%
\providecommand \BibitemShut  [1]{\csname bibitem#1\endcsname}%
\let\auto@bib@innerbib\@empty
\bibitem [{\citenamefont {Aharonov}\ \emph {et~al.}(1988)\citenamefont
  {Aharonov}, \citenamefont {Albert},\ and\ \citenamefont
  {Vaidman}}]{aharonov88}%
  \BibitemOpen
  \bibfield  {author} {\bibinfo {author} {\bibfnamefont {Y.}~\bibnamefont
  {Aharonov}}, \bibinfo {author} {\bibfnamefont {D.~Z.}\ \bibnamefont
  {Albert}}, \ and\ \bibinfo {author} {\bibfnamefont {L.}~\bibnamefont
  {Vaidman}},\ }\bibfield  {title} {\enquote {\bibinfo {title} {How the result
  of a measurement of a component of the spin of a spin-1/2 particle can turn
  out to be 100},}\ }\href {\doibase 10.1103/PhysRevLett.60.1351} {\bibfield
  {journal} {\bibinfo  {journal} {Phys. Rev. Lett.}\ }\textbf {\bibinfo
  {volume} {60}},\ \bibinfo {pages} {1351} (\bibinfo {year}
  {1988})}\BibitemShut {NoStop}%
\bibitem [{\citenamefont {Hosten}\ and\ \citenamefont
  {Kwiat}(2008)}]{hosten08}%
  \BibitemOpen
  \bibfield  {author} {\bibinfo {author} {\bibfnamefont {O.}~\bibnamefont
  {Hosten}}\ and\ \bibinfo {author} {\bibfnamefont {P.}~\bibnamefont {Kwiat}},\
  }\bibfield  {title} {\enquote {\bibinfo {title} {Observation of the spin
  {H}all effect of light via weak measurements},}\ }\href {\doibase
  10.1126/science.1152697} {\bibfield  {journal} {\bibinfo  {journal}
  {Science}\ }\textbf {\bibinfo {volume} {319}},\ \bibinfo {pages} {787}
  (\bibinfo {year} {2008})}\BibitemShut {NoStop}%
\bibitem [{\citenamefont {Dixon}\ \emph {et~al.}(2009)\citenamefont {Dixon},
  \citenamefont {Starling}, \citenamefont {Jordan},\ and\ \citenamefont
  {Howell}}]{dixon09}%
  \BibitemOpen
  \bibfield  {author} {\bibinfo {author} {\bibfnamefont {P.~B.}\ \bibnamefont
  {Dixon}}, \bibinfo {author} {\bibfnamefont {D.~J.}\ \bibnamefont {Starling}},
  \bibinfo {author} {\bibfnamefont {A.~N.}\ \bibnamefont {Jordan}}, \ and\
  \bibinfo {author} {\bibfnamefont {J.~C.}\ \bibnamefont {Howell}},\ }\bibfield
   {title} {\enquote {\bibinfo {title} {Ultrasensitive beam deflection
  measurement via interferometric weak value amplification},}\ }\href {\doibase
  10.1103/PhysRevLett.102.173601} {\bibfield  {journal} {\bibinfo  {journal}
  {Phys. Rev. Lett.}\ }\textbf {\bibinfo {volume} {102}},\ \bibinfo {pages}
  {173601} (\bibinfo {year} {2009})}\BibitemShut {NoStop}%
\bibitem [{\citenamefont {Alves}\ \emph {et~al.}(2015)\citenamefont {Alves},
  \citenamefont {Escher}, \citenamefont {de~Matos~Filho}, \citenamefont
  {Zagury},\ and\ \citenamefont {Davidovich}}]{alves15}%
  \BibitemOpen
  \bibfield  {author} {\bibinfo {author} {\bibfnamefont {G.~B.}\ \bibnamefont
  {Alves}}, \bibinfo {author} {\bibfnamefont {B.~M.}\ \bibnamefont {Escher}},
  \bibinfo {author} {\bibfnamefont {R.~L.}\ \bibnamefont {de~Matos~Filho}},
  \bibinfo {author} {\bibfnamefont {N.}~\bibnamefont {Zagury}}, \ and\ \bibinfo
  {author} {\bibfnamefont {L.}~\bibnamefont {Davidovich}},\ }\bibfield  {title}
  {\enquote {\bibinfo {title} {Weak-value amplification as an optimal
  metrological protocol},}\ }\href {\doibase 10.1103/PhysRevA.91.062107}
  {\bibfield  {journal} {\bibinfo  {journal} {Phys. Rev. A}\ }\textbf {\bibinfo
  {volume} {91}},\ \bibinfo {pages} {062107} (\bibinfo {year}
  {2015})}\BibitemShut {NoStop}%
\bibitem [{\citenamefont {Kim}\ \emph {et~al.}(2022)\citenamefont {Kim},
  \citenamefont {Yoo},\ and\ \citenamefont {Kim}}]{kim22}%
  \BibitemOpen
  \bibfield  {author} {\bibinfo {author} {\bibfnamefont {Y.}~\bibnamefont
  {Kim}}, \bibinfo {author} {\bibfnamefont {S.-Y.}\ \bibnamefont {Yoo}}, \ and\
  \bibinfo {author} {\bibfnamefont {Y.-H.}\ \bibnamefont {Kim}},\ }\bibfield
  {title} {\enquote {\bibinfo {title} {Heisenberg-limited metrology via
  weak-value amplification without using entangled resources},}\ }\href
  {\doibase 10.1103/PhysRevLett.128.040503} {\bibfield  {journal} {\bibinfo
  {journal} {Phys. Rev. Lett.}\ }\textbf {\bibinfo {volume} {128}},\ \bibinfo
  {pages} {040503} (\bibinfo {year} {2022})}\BibitemShut {NoStop}%
\bibitem [{\citenamefont {Lundeen}\ \emph {et~al.}(2011)\citenamefont
  {Lundeen}, \citenamefont {Sutherland}, \citenamefont {Patel}, \citenamefont
  {Stewart},\ and\ \citenamefont {Bamber}}]{lundeen11}%
  \BibitemOpen
  \bibfield  {author} {\bibinfo {author} {\bibfnamefont {J.~S.}\ \bibnamefont
  {Lundeen}}, \bibinfo {author} {\bibfnamefont {B.}~\bibnamefont {Sutherland}},
  \bibinfo {author} {\bibfnamefont {A.}~\bibnamefont {Patel}}, \bibinfo
  {author} {\bibfnamefont {C.}~\bibnamefont {Stewart}}, \ and\ \bibinfo
  {author} {\bibfnamefont {C.}~\bibnamefont {Bamber}},\ }\bibfield  {title}
  {\enquote {\bibinfo {title} {Direct measurement of the quantum
  wavefunction},}\ }\href {\doibase 10.1038/nature10120} {\bibfield  {journal}
  {\bibinfo  {journal} {Nature}\ }\textbf {\bibinfo {volume} {474}},\ \bibinfo
  {pages} {188} (\bibinfo {year} {2011})}\BibitemShut {NoStop}%
\bibitem [{\citenamefont {Mirhosseini}\ \emph {et~al.}(2014)\citenamefont
  {Mirhosseini}, \citenamefont {Maga\~na Loaiza}, \citenamefont {Rafsanjani},\
  and\ \citenamefont {Boyd}}]{mirhosseini14}%
  \BibitemOpen
  \bibfield  {author} {\bibinfo {author} {\bibfnamefont {M.}~\bibnamefont
  {Mirhosseini}}, \bibinfo {author} {\bibfnamefont {O.~S.}\ \bibnamefont
  {Maga\~na Loaiza}}, \bibinfo {author} {\bibfnamefont {S.~M.~H.}\ \bibnamefont
  {Rafsanjani}}, \ and\ \bibinfo {author} {\bibfnamefont {R.~W.}\ \bibnamefont
  {Boyd}},\ }\bibfield  {title} {\enquote {\bibinfo {title} {Compressive direct
  measurement of the quantum wave function},}\ }\href {\doibase
  10.1103/PhysRevLett.113.090402} {\bibfield  {journal} {\bibinfo  {journal}
  {Phys. Rev. Lett.}\ }\textbf {\bibinfo {volume} {113}},\ \bibinfo {pages}
  {090402} (\bibinfo {year} {2014})}\BibitemShut {NoStop}%
\bibitem [{\citenamefont {Pan}\ \emph {et~al.}(2019)\citenamefont {Pan},
  \citenamefont {Xu}, \citenamefont {Kedem}, \citenamefont {Wang},
  \citenamefont {Chen}, \citenamefont {Jan}, \citenamefont {Sun}, \citenamefont
  {Xu}, \citenamefont {Han}, \citenamefont {Li},\ and\ \citenamefont
  {Guo}}]{pan19}%
  \BibitemOpen
  \bibfield  {author} {\bibinfo {author} {\bibfnamefont {W.-W.}\ \bibnamefont
  {Pan}}, \bibinfo {author} {\bibfnamefont {X.-Y.}\ \bibnamefont {Xu}},
  \bibinfo {author} {\bibfnamefont {Y.}~\bibnamefont {Kedem}}, \bibinfo
  {author} {\bibfnamefont {Q.-Q.}\ \bibnamefont {Wang}}, \bibinfo {author}
  {\bibfnamefont {Z.}~\bibnamefont {Chen}}, \bibinfo {author} {\bibfnamefont
  {M.}~\bibnamefont {Jan}}, \bibinfo {author} {\bibfnamefont {K.}~\bibnamefont
  {Sun}}, \bibinfo {author} {\bibfnamefont {J.-S.}\ \bibnamefont {Xu}},
  \bibinfo {author} {\bibfnamefont {Y.-J.}\ \bibnamefont {Han}}, \bibinfo
  {author} {\bibfnamefont {C.-F.}\ \bibnamefont {Li}}, \ and\ \bibinfo {author}
  {\bibfnamefont {G.-C.}\ \bibnamefont {Guo}},\ }\bibfield  {title} {\enquote
  {\bibinfo {title} {Direct measurement of a nonlocal entangled quantum
  state},}\ }\href {\doibase 10.1103/PhysRevLett.123.150402} {\bibfield
  {journal} {\bibinfo  {journal} {Phys. Rev. Lett.}\ }\textbf {\bibinfo
  {volume} {123}},\ \bibinfo {pages} {150402} (\bibinfo {year}
  {2019})}\BibitemShut {NoStop}%
\bibitem [{\citenamefont {Sj{\"o}qvist}(2006)}]{sjoqvist06}%
  \BibitemOpen
  \bibfield  {author} {\bibinfo {author} {\bibfnamefont {E.}~\bibnamefont
  {Sj{\"o}qvist}},\ }\bibfield  {title} {\enquote {\bibinfo {title} {Geometric
  phase in weak measurements},}\ }\href {\doibase
  10.1016/j.physleta.2006.06.028} {\bibfield  {journal} {\bibinfo  {journal}
  {Phys. Lett. A}\ }\textbf {\bibinfo {volume} {359}},\ \bibinfo {pages} {187}
  (\bibinfo {year} {2006})}\BibitemShut {NoStop}%
\bibitem [{\citenamefont {Cho}\ \emph {et~al.}(2019)\citenamefont {Cho},
  \citenamefont {Kim}, \citenamefont {Choi}, \citenamefont {Kim}, \citenamefont
  {Han}, \citenamefont {Lee}, \citenamefont {Moon},\ and\ \citenamefont
  {Kim}}]{cho19}%
  \BibitemOpen
  \bibfield  {author} {\bibinfo {author} {\bibfnamefont {Y.-W.}\ \bibnamefont
  {Cho}}, \bibinfo {author} {\bibfnamefont {Y.}~\bibnamefont {Kim}}, \bibinfo
  {author} {\bibfnamefont {Y.-H.}\ \bibnamefont {Choi}}, \bibinfo {author}
  {\bibfnamefont {Y.-S.}\ \bibnamefont {Kim}}, \bibinfo {author} {\bibfnamefont
  {S.-W.}\ \bibnamefont {Han}}, \bibinfo {author} {\bibfnamefont {S.-Y.}\
  \bibnamefont {Lee}}, \bibinfo {author} {\bibfnamefont {S.}~\bibnamefont
  {Moon}}, \ and\ \bibinfo {author} {\bibfnamefont {Y.-Ho.}\ \bibnamefont
  {Kim}},\ }\bibfield  {title} {\enquote {\bibinfo {title} {Emergence of the
  geometric phase from quantum measurement back-action},}\ }\href {\doibase
  10.1038/s41567-019-0482-z} {\bibfield  {journal} {\bibinfo  {journal} {Nat.
  Phys.}\ }\textbf {\bibinfo {volume} {15}},\ \bibinfo {pages} {665} (\bibinfo
  {year} {2019})}\BibitemShut {NoStop}%
\bibitem [{\citenamefont {Dressel}\ \emph {et~al.}(2014)\citenamefont
  {Dressel}, \citenamefont {Malik}, \citenamefont {Miatto}, \citenamefont
  {Jordan},\ and\ \citenamefont {Boyd}}]{dressel14}%
  \BibitemOpen
  \bibfield  {author} {\bibinfo {author} {\bibfnamefont {J.}~\bibnamefont
  {Dressel}}, \bibinfo {author} {\bibfnamefont {M.}~\bibnamefont {Malik}},
  \bibinfo {author} {\bibfnamefont {F.~M.}\ \bibnamefont {Miatto}}, \bibinfo
  {author} {\bibfnamefont {A.~N.}\ \bibnamefont {Jordan}}, \ and\ \bibinfo
  {author} {\bibfnamefont {R.~W.}\ \bibnamefont {Boyd}},\ }\bibfield  {title}
  {\enquote {\bibinfo {title} {Colloquium: Understanding quantum weak values:
  Basics and applications},}\ }\href {\doibase 10.1103/RevModPhys.86.307}
  {\bibfield  {journal} {\bibinfo  {journal} {Rev. Mod. Phys.}\ }\textbf
  {\bibinfo {volume} {86}},\ \bibinfo {pages} {307} (\bibinfo {year}
  {2014})}\BibitemShut {NoStop}%
\bibitem [{\citenamefont {Huang}\ \emph {et~al.}(2019)\citenamefont {Huang},
  \citenamefont {Lee}, \citenamefont {Zhang}, \citenamefont {Fei},\ and\
  \citenamefont {Wu}}]{huang19}%
  \BibitemOpen
  \bibfield  {author} {\bibinfo {author} {\bibfnamefont {M.}~\bibnamefont
  {Huang}}, \bibinfo {author} {\bibfnamefont {R.-K.}\ \bibnamefont {Lee}},
  \bibinfo {author} {\bibfnamefont {L.}~\bibnamefont {Zhang}}, \bibinfo
  {author} {\bibfnamefont {S.-M.}\ \bibnamefont {Fei}}, \ and\ \bibinfo
  {author} {\bibfnamefont {J.}~\bibnamefont {Wu}},\ }\bibfield  {title}
  {\enquote {\bibinfo {title} {Simulating broken $\mathcal{PT}$-symmetric
  {H}amiltonian systems by weak measurement},}\ }\href {\doibase
  10.1103/PhysRevLett.123.080404} {\bibfield  {journal} {\bibinfo  {journal}
  {Phys. Rev. Lett.}\ }\textbf {\bibinfo {volume} {123}},\ \bibinfo {pages}
  {080404} (\bibinfo {year} {2019})}\BibitemShut {NoStop}%
\bibitem [{\citenamefont {Wagner}\ \emph {et~al.}(2021)\citenamefont {Wagner},
  \citenamefont {Kersten}, \citenamefont {Danner}, \citenamefont {Lemmel},
  \citenamefont {Pan},\ and\ \citenamefont {Sponar}}]{wagner21}%
  \BibitemOpen
  \bibfield  {author} {\bibinfo {author} {\bibfnamefont {R.}~\bibnamefont
  {Wagner}}, \bibinfo {author} {\bibfnamefont {W.}~\bibnamefont {Kersten}},
  \bibinfo {author} {\bibfnamefont {A.}~\bibnamefont {Danner}}, \bibinfo
  {author} {\bibfnamefont {H.}~\bibnamefont {Lemmel}}, \bibinfo {author}
  {\bibfnamefont {A.~K.}\ \bibnamefont {Pan}}, \ and\ \bibinfo {author}
  {\bibfnamefont {S.}~\bibnamefont {Sponar}},\ }\bibfield  {title} {\enquote
  {\bibinfo {title} {Direct experimental test of commutation relation via
  imaginary weak value},}\ }\href {\doibase 10.1103/PhysRevResearch.3.023243}
  {\bibfield  {journal} {\bibinfo  {journal} {Phys. Rev. Res.}\ }\textbf
  {\bibinfo {volume} {3}},\ \bibinfo {pages} {023243} (\bibinfo {year}
  {2021})}\BibitemShut {NoStop}%
\bibitem [{\citenamefont {Aharonov}\ and\ \citenamefont
  {Vaidman}(1991)}]{aharonov91}%
  \BibitemOpen
  \bibfield  {author} {\bibinfo {author} {\bibfnamefont {Y.}~\bibnamefont
  {Aharonov}}\ and\ \bibinfo {author} {\bibfnamefont {L.}~\bibnamefont
  {Vaidman}},\ }\bibfield  {title} {\enquote {\bibinfo {title} {Complete
  description of a quantum system at a given time},}\ }\href {\doibase
  10.1088/0305-4470/24/10/018} {\bibfield  {journal} {\bibinfo  {journal} {J.
  Phys. A Math. Gen.}\ }\textbf {\bibinfo {volume} {24}},\ \bibinfo {pages}
  {2315} (\bibinfo {year} {1991})}\BibitemShut {NoStop}%
\bibitem [{\citenamefont {Resch}\ \emph {et~al.}(2004)\citenamefont {Resch},
  \citenamefont {Lundeen},\ and\ \citenamefont {Steinberg}}]{resch04}%
  \BibitemOpen
  \bibfield  {author} {\bibinfo {author} {\bibfnamefont {K.~J.}\ \bibnamefont
  {Resch}}, \bibinfo {author} {\bibfnamefont {J.~S.}\ \bibnamefont {Lundeen}},
  \ and\ \bibinfo {author} {\bibfnamefont {A.~M.}\ \bibnamefont {Steinberg}},\
  }\bibfield  {title} {\enquote {\bibinfo {title} {Experimental realization of
  the quantum box problem},}\ }\href {\doibase 10.1016/j.physleta.2004.02.042}
  {\bibfield  {journal} {\bibinfo  {journal} {Phys. Lett. A}\ }\textbf
  {\bibinfo {volume} {324}},\ \bibinfo {pages} {125} (\bibinfo {year}
  {2004})}\BibitemShut {NoStop}%
\bibitem [{\citenamefont {Vaidman}(2013)}]{vaidman13}%
  \BibitemOpen
  \bibfield  {author} {\bibinfo {author} {\bibfnamefont {L.}~\bibnamefont
  {Vaidman}},\ }\bibfield  {title} {\enquote {\bibinfo {title} {Past of a
  quantum particle},}\ }\href {\doibase 10.1103/PhysRevA.87.052104} {\bibfield
  {journal} {\bibinfo  {journal} {Phys. Rev. A}\ }\textbf {\bibinfo {volume}
  {87}},\ \bibinfo {pages} {052104} (\bibinfo {year} {2013})}\BibitemShut
  {NoStop}%
\bibitem [{\citenamefont {Danan}\ \emph {et~al.}(2013)\citenamefont {Danan},
  \citenamefont {Farfurnik}, \citenamefont {Bar-Ad},\ and\ \citenamefont
  {Vaidman}}]{danan13}%
  \BibitemOpen
  \bibfield  {author} {\bibinfo {author} {\bibfnamefont {A.}~\bibnamefont
  {Danan}}, \bibinfo {author} {\bibfnamefont {D.}~\bibnamefont {Farfurnik}},
  \bibinfo {author} {\bibfnamefont {S.}~\bibnamefont {Bar-Ad}}, \ and\ \bibinfo
  {author} {\bibfnamefont {L.}~\bibnamefont {Vaidman}},\ }\bibfield  {title}
  {\enquote {\bibinfo {title} {Asking photons where they have been},}\ }\href
  {\doibase 10.1103/PhysRevLett.111.240402} {\bibfield  {journal} {\bibinfo
  {journal} {Phys. Rev. Lett.}\ }\textbf {\bibinfo {volume} {111}},\ \bibinfo
  {pages} {240402} (\bibinfo {year} {2013})}\BibitemShut {NoStop}%
\bibitem [{\citenamefont {Aharonov}\ \emph {et~al.}(2013)\citenamefont
  {Aharonov}, \citenamefont {Popescu}, \citenamefont {Rohrlich},\ and\
  \citenamefont {Skrzypczyk}}]{aharonov13}%
  \BibitemOpen
  \bibfield  {author} {\bibinfo {author} {\bibfnamefont {Y.}~\bibnamefont
  {Aharonov}}, \bibinfo {author} {\bibfnamefont {S.}~\bibnamefont {Popescu}},
  \bibinfo {author} {\bibfnamefont {D.}~\bibnamefont {Rohrlich}}, \ and\
  \bibinfo {author} {\bibfnamefont {P.}~\bibnamefont {Skrzypczyk}},\ }\bibfield
   {title} {\enquote {\bibinfo {title} {Quantum {C}heshire cats},}\ }\href
  {\doibase 10.1088/1367-2630/15/11/113015} {\bibfield  {journal} {\bibinfo
  {journal} {New J. Phys.}\ }\textbf {\bibinfo {volume} {15}},\ \bibinfo
  {pages} {113015} (\bibinfo {year} {2013})}\BibitemShut {NoStop}%
\bibitem [{\citenamefont {Denkmayr}\ \emph {et~al.}(2014)\citenamefont
  {Denkmayr}, \citenamefont {Geppert}, \citenamefont {Sponar}, \citenamefont
  {Lemmel}, \citenamefont {Matzkin}, \citenamefont {Tollaksen},\ and\
  \citenamefont {Hasegawa}}]{denkmayr14}%
  \BibitemOpen
  \bibfield  {author} {\bibinfo {author} {\bibfnamefont {T.}~\bibnamefont
  {Denkmayr}}, \bibinfo {author} {\bibfnamefont {H.}~\bibnamefont {Geppert}},
  \bibinfo {author} {\bibfnamefont {S.}~\bibnamefont {Sponar}}, \bibinfo
  {author} {\bibfnamefont {H.}~\bibnamefont {Lemmel}}, \bibinfo {author}
  {\bibfnamefont {A.}~\bibnamefont {Matzkin}}, \bibinfo {author} {\bibfnamefont
  {J.}~\bibnamefont {Tollaksen}}, \ and\ \bibinfo {author} {\bibfnamefont
  {Y.}~\bibnamefont {Hasegawa}},\ }\bibfield  {title} {\enquote {\bibinfo
  {title} {Observation of a quantum {C}heshire cat in a matter-wave
  interferometer experiment},}\ }\href {\doibase 10.1038/ncomms5492} {\bibfield
   {journal} {\bibinfo  {journal} {Nat. Commun.}\ }\textbf {\bibinfo {volume}
  {5}},\ \bibinfo {pages} {4492} (\bibinfo {year} {2014})}\BibitemShut
  {NoStop}%
\bibitem [{\citenamefont {Das}\ and\ \citenamefont {Pati}(2020)}]{das20}%
  \BibitemOpen
  \bibfield  {author} {\bibinfo {author} {\bibfnamefont {D.}~\bibnamefont
  {Das}}\ and\ \bibinfo {author} {\bibfnamefont {A.~K.}\ \bibnamefont {Pati}},\
  }\bibfield  {title} {\enquote {\bibinfo {title} {Can two quantum {C}heshire
  cats exchange grins?}}\ }\href {\doibase 10.1088/1367-2630/ab8e5a} {\bibfield
   {journal} {\bibinfo  {journal} {New J. Phys.}\ }\textbf {\bibinfo {volume}
  {22}},\ \bibinfo {pages} {063032} (\bibinfo {year} {2020})}\BibitemShut
  {NoStop}%
\bibitem [{\citenamefont {\textit{et al.}}(2020)}]{liu20}%
  \BibitemOpen
  \bibfield  {author} {\bibinfo {author} {\bibfnamefont {Z.-H.~Liu}\
  \bibnamefont {\textit{et al.}}},\ }\bibfield  {title} {\enquote {\bibinfo
  {title} {Experimental exchange of grins between quantum {C}heshire cats},}\
  }\href {\doibase 10.1038/s41467-020-16761-0} {\bibfield  {journal} {\bibinfo
  {journal} {Nat. Commun.}\ }\textbf {\bibinfo {volume} {11}},\ \bibinfo
  {pages} {3006} (\bibinfo {year} {2020})}\BibitemShut {NoStop}%
\bibitem [{\citenamefont {Aharonov}\ \emph {et~al.}(2016)\citenamefont
  {Aharonov}, \citenamefont {Colombo}, \citenamefont {Popescu}, \citenamefont
  {Sabadini}, \citenamefont {Struppa},\ and\ \citenamefont
  {Tollaksen}}]{aharonov16}%
  \BibitemOpen
  \bibfield  {author} {\bibinfo {author} {\bibfnamefont {Y.}~\bibnamefont
  {Aharonov}}, \bibinfo {author} {\bibfnamefont {F.}~\bibnamefont {Colombo}},
  \bibinfo {author} {\bibfnamefont {S.}~\bibnamefont {Popescu}}, \bibinfo
  {author} {\bibfnamefont {I.}~\bibnamefont {Sabadini}}, \bibinfo {author}
  {\bibfnamefont {D.~C.}\ \bibnamefont {Struppa}}, \ and\ \bibinfo {author}
  {\bibfnamefont {J.}~\bibnamefont {Tollaksen}},\ }\bibfield  {title} {\enquote
  {\bibinfo {title} {Quantum violation of the pigeonhole principle and the
  nature of quantum correlations},}\ }\href {\doibase 10.1073/pnas.1522411112}
  {\bibfield  {journal} {\bibinfo  {journal} {Proc. Natl. Acad. Sci. USA}\
  }\textbf {\bibinfo {volume} {113}},\ \bibinfo {pages} {532} (\bibinfo {year}
  {2016})}\BibitemShut {NoStop}%
\bibitem [{\citenamefont {Waegell}\ \emph {et~al.}(2017)\citenamefont
  {Waegell}, \citenamefont {Denkmayr}, \citenamefont {Geppert}, \citenamefont
  {Ebner}, \citenamefont {Jenke}, \citenamefont {Hasegawa}, \citenamefont
  {Sponar}, \citenamefont {Dressel},\ and\ \citenamefont
  {Tollaksen}}]{waegell17}%
  \BibitemOpen
  \bibfield  {author} {\bibinfo {author} {\bibfnamefont {M.}~\bibnamefont
  {Waegell}}, \bibinfo {author} {\bibfnamefont {T.}~\bibnamefont {Denkmayr}},
  \bibinfo {author} {\bibfnamefont {H.}~\bibnamefont {Geppert}}, \bibinfo
  {author} {\bibfnamefont {D.}~\bibnamefont {Ebner}}, \bibinfo {author}
  {\bibfnamefont {T.}~\bibnamefont {Jenke}}, \bibinfo {author} {\bibfnamefont
  {Y.}~\bibnamefont {Hasegawa}}, \bibinfo {author} {\bibfnamefont
  {S.}~\bibnamefont {Sponar}}, \bibinfo {author} {\bibfnamefont
  {J.}~\bibnamefont {Dressel}}, \ and\ \bibinfo {author} {\bibfnamefont
  {J.}~\bibnamefont {Tollaksen}},\ }\bibfield  {title} {\enquote {\bibinfo
  {title} {Confined contextuality in neutron interferometry: Observing the
  quantum pigeonhole effect},}\ }\href {\doibase 10.1103/PhysRevA.96.052131}
  {\bibfield  {journal} {\bibinfo  {journal} {Phys. Rev. A}\ }\textbf {\bibinfo
  {volume} {96}},\ \bibinfo {pages} {052131} (\bibinfo {year}
  {2017})}\BibitemShut {NoStop}%
\bibitem [{\citenamefont {\textit{et al.}}(2019)}]{chen19}%
  \BibitemOpen
  \bibfield  {author} {\bibinfo {author} {\bibfnamefont {M.-C.~Chen}\
  \bibnamefont {\textit{et al.}}},\ }\bibfield  {title} {\enquote {\bibinfo
  {title} {Experimental demonstration of quantum pigeonhole paradox},}\ }\href
  {\doibase 10.1073/pnas.1815462116} {\bibfield  {journal} {\bibinfo  {journal}
  {Proc. Natl. Acad. Sci. USA}\ }\textbf {\bibinfo {volume} {116}},\ \bibinfo
  {pages} {1549} (\bibinfo {year} {2019})}\BibitemShut {NoStop}%
\bibitem [{\citenamefont {Reznik}\ \emph {et~al.}(2020)\citenamefont {Reznik},
  \citenamefont {Bagchi}, \citenamefont {Dressel},\ and\ \citenamefont
  {Vaidman}}]{reznik20}%
  \BibitemOpen
  \bibfield  {author} {\bibinfo {author} {\bibfnamefont {G.}~\bibnamefont
  {Reznik}}, \bibinfo {author} {\bibfnamefont {S.}~\bibnamefont {Bagchi}},
  \bibinfo {author} {\bibfnamefont {J.}~\bibnamefont {Dressel}}, \ and\
  \bibinfo {author} {\bibfnamefont {L.}~\bibnamefont {Vaidman}},\ }\bibfield
  {title} {\enquote {\bibinfo {title} {Footprints of quantum pigeons},}\ }\href
  {\doibase 10.1103/PhysRevResearch.2.023004} {\bibfield  {journal} {\bibinfo
  {journal} {Phys. Rev. Research}\ }\textbf {\bibinfo {volume} {2}},\ \bibinfo
  {pages} {023004} (\bibinfo {year} {2020})}\BibitemShut {NoStop}%
\bibitem [{\citenamefont {Duck}\ \emph {et~al.}(1989)\citenamefont {Duck},
  \citenamefont {Stevenson},\ and\ \citenamefont {Sudarshan}}]{duck89}%
  \BibitemOpen
  \bibfield  {author} {\bibinfo {author} {\bibfnamefont {I.~M.}\ \bibnamefont
  {Duck}}, \bibinfo {author} {\bibfnamefont {P.~M.}\ \bibnamefont {Stevenson}},
  \ and\ \bibinfo {author} {\bibfnamefont {E.~C.~G.}\ \bibnamefont
  {Sudarshan}},\ }\bibfield  {title} {\enquote {\bibinfo {title} {The sense in
  which a ``weak measurement'' of a spin-$1/2$ particle's spin component yields
  a value 100},}\ }\href {\doibase 10.1103/PhysRevD.40.2112} {\bibfield
  {journal} {\bibinfo  {journal} {Phys. Rev. D}\ }\textbf {\bibinfo {volume}
  {40}},\ \bibinfo {pages} {2112} (\bibinfo {year} {1989})}\BibitemShut
  {NoStop}%
\bibitem [{\citenamefont {Saldanha}(2014)}]{saldanha14}%
  \BibitemOpen
  \bibfield  {author} {\bibinfo {author} {\bibfnamefont {P.~L.}\ \bibnamefont
  {Saldanha}},\ }\bibfield  {title} {\enquote {\bibinfo {title} {Interpreting a
  nested {M}ach-{Z}ehnder interferometer with classical optics},}\ }\href
  {\doibase 10.1103/PhysRevA.89.033825} {\bibfield  {journal} {\bibinfo
  {journal} {Phys. Rev. A}\ }\textbf {\bibinfo {volume} {89}},\ \bibinfo
  {pages} {033825} (\bibinfo {year} {2014})}\BibitemShut {NoStop}%
\bibitem [{\citenamefont {Bartkiewicz}\ \emph {et~al.}(2015)\citenamefont
  {Bartkiewicz}, \citenamefont {\ifmmode~\check{C}\else \v{C}\fi{}ernoch},
  \citenamefont {Jav\ifmmode~\mathring{u}\else \r{u}\fi{}rek}, \citenamefont
  {Lemr}, \citenamefont {Soubusta},\ and\ \citenamefont
  {Svozil\'{\i}k}}]{bartkiewicz15}%
  \BibitemOpen
  \bibfield  {author} {\bibinfo {author} {\bibfnamefont {K.}~\bibnamefont
  {Bartkiewicz}}, \bibinfo {author} {\bibfnamefont {A.}~\bibnamefont
  {\ifmmode~\check{C}\else \v{C}\fi{}ernoch}}, \bibinfo {author} {\bibfnamefont
  {D.}~\bibnamefont {Jav\ifmmode~\mathring{u}\else \r{u}\fi{}rek}}, \bibinfo
  {author} {\bibfnamefont {K.}~\bibnamefont {Lemr}}, \bibinfo {author}
  {\bibfnamefont {J.}~\bibnamefont {Soubusta}}, \ and\ \bibinfo {author}
  {\bibfnamefont {J.}~\bibnamefont {Svozil\'{\i}k}},\ }\bibfield  {title}
  {\enquote {\bibinfo {title} {One-state vector formalism for the evolution of
  a quantum state through nested {M}ach-{Z}ehnder interferometers},}\ }\href
  {\doibase 10.1103/PhysRevA.91.012103} {\bibfield  {journal} {\bibinfo
  {journal} {Phys. Rev. A}\ }\textbf {\bibinfo {volume} {91}},\ \bibinfo
  {pages} {012103} (\bibinfo {year} {2015})}\BibitemShut {NoStop}%
\bibitem [{\citenamefont {Englert}\ \emph {et~al.}(2017)\citenamefont
  {Englert}, \citenamefont {Horia}, \citenamefont {Dai}, \citenamefont {Len},\
  and\ \citenamefont {Ng}}]{englert17}%
  \BibitemOpen
  \bibfield  {author} {\bibinfo {author} {\bibfnamefont {B.-G.}\ \bibnamefont
  {Englert}}, \bibinfo {author} {\bibfnamefont {K.}~\bibnamefont {Horia}},
  \bibinfo {author} {\bibfnamefont {J.}~\bibnamefont {Dai}}, \bibinfo {author}
  {\bibfnamefont {Y.~L.}\ \bibnamefont {Len}}, \ and\ \bibinfo {author}
  {\bibfnamefont {H.~K.}\ \bibnamefont {Ng}},\ }\bibfield  {title} {\enquote
  {\bibinfo {title} {Past of a quantum particle revisited},}\ }\href {\doibase
  10.1103/PhysRevA.96.022126} {\bibfield  {journal} {\bibinfo  {journal} {Phys.
  Rev. A}\ }\textbf {\bibinfo {volume} {96}},\ \bibinfo {pages} {022126}
  (\bibinfo {year} {2017})}\BibitemShut {NoStop}%
\bibitem [{\citenamefont {Corr{\^{e}}a}\ \emph {et~al.}(2015)\citenamefont
  {Corr{\^{e}}a}, \citenamefont {Santos}, \citenamefont {Monken},\ and\
  \citenamefont {Saldanha}}]{correa15}%
  \BibitemOpen
  \bibfield  {author} {\bibinfo {author} {\bibfnamefont {R.}~\bibnamefont
  {Corr{\^{e}}a}}, \bibinfo {author} {\bibfnamefont {M.~F.}\ \bibnamefont
  {Santos}}, \bibinfo {author} {\bibfnamefont {C.~H.}\ \bibnamefont {Monken}},
  \ and\ \bibinfo {author} {\bibfnamefont {P.~L.}\ \bibnamefont {Saldanha}},\
  }\bibfield  {title} {\enquote {\bibinfo {title} {`{Q}uantum {C}heshire cat'
  as simple quantum interference},}\ }\href {\doibase
  10.1088/1367-2630/17/5/053042} {\bibfield  {journal} {\bibinfo  {journal}
  {New J. Phys.}\ }\textbf {\bibinfo {volume} {17}},\ \bibinfo {pages} {053042}
  (\bibinfo {year} {2015})}\BibitemShut {NoStop}%
\bibitem [{\citenamefont {Atherton}\ \emph {et~al.}(2015)\citenamefont
  {Atherton}, \citenamefont {Ranjit}, \citenamefont {Geraci},\ and\
  \citenamefont {Weinstein}}]{atherton15}%
  \BibitemOpen
  \bibfield  {author} {\bibinfo {author} {\bibfnamefont {D.~P.}\ \bibnamefont
  {Atherton}}, \bibinfo {author} {\bibfnamefont {G.}~\bibnamefont {Ranjit}},
  \bibinfo {author} {\bibfnamefont {A.~A.}\ \bibnamefont {Geraci}}, \ and\
  \bibinfo {author} {\bibfnamefont {J.~D.}\ \bibnamefont {Weinstein}},\
  }\bibfield  {title} {\enquote {\bibinfo {title} {Observation of a classical
  {C}heshire cat in an optical interferometer},}\ }\href {\doibase
  10.1364/OL.40.000879} {\bibfield  {journal} {\bibinfo  {journal} {Opt.
  Lett.}\ }\textbf {\bibinfo {volume} {40}},\ \bibinfo {pages} {879} (\bibinfo
  {year} {2015})}\BibitemShut {NoStop}%
\bibitem [{\citenamefont {Corr\^ea}\ and\ \citenamefont
  {Saldanha}(2021)}]{correa21}%
  \BibitemOpen
  \bibfield  {author} {\bibinfo {author} {\bibfnamefont {R.}~\bibnamefont
  {Corr\^ea}}\ and\ \bibinfo {author} {\bibfnamefont {P.~L.}\ \bibnamefont
  {Saldanha}},\ }\bibfield  {title} {\enquote {\bibinfo {title} {Apparent
  quantum paradoxes as simple interference: Quantum violation of the pigeonhole
  principle and exchange of properties between quantum particles},}\ }\href
  {\doibase 10.1103/PhysRevA.104.012212} {\bibfield  {journal} {\bibinfo
  {journal} {Phys. Rev. A}\ }\textbf {\bibinfo {volume} {104}},\ \bibinfo
  {pages} {012212} (\bibinfo {year} {2021})}\BibitemShut {NoStop}%
\bibitem [{\citenamefont {Saldanha}(2020)}]{saldanha20}%
  \BibitemOpen
  \bibfield  {author} {\bibinfo {author} {\bibfnamefont {P.~L.}\ \bibnamefont
  {Saldanha}},\ }\bibfield  {title} {\enquote {\bibinfo {title} {Inconsistency
  of a realistic interpretation of quantum measurements: a simple example},}\
  }\href {\doibase 10.1007/s13538-020-00757-8} {\bibfield  {journal} {\bibinfo
  {journal} {Braz. J. Phys.}\ }\textbf {\bibinfo {volume} {50}},\ \bibinfo
  {pages} {438} (\bibinfo {year} {2020})}\BibitemShut {NoStop}%
\bibitem [{\citenamefont {Budroni}\ \emph {et~al.}(2022)\citenamefont
  {Budroni}, \citenamefont {Cabello}, \citenamefont {G\"uhne}, \citenamefont
  {Kleinmann},\ and\ \citenamefont {Larsson}}]{budroni22}%
  \BibitemOpen
  \bibfield  {author} {\bibinfo {author} {\bibfnamefont {C.}~\bibnamefont
  {Budroni}}, \bibinfo {author} {\bibfnamefont {A.}~\bibnamefont {Cabello}},
  \bibinfo {author} {\bibfnamefont {O.}~\bibnamefont {G\"uhne}}, \bibinfo
  {author} {\bibfnamefont {M.}~\bibnamefont {Kleinmann}}, \ and\ \bibinfo
  {author} {\bibfnamefont {J.-\AA{}.}\ \bibnamefont {Larsson}},\ }\bibfield
  {title} {\enquote {\bibinfo {title} {Kochen-{S}pecker contextuality},}\
  }\href {\doibase 10.1103/RevModPhys.94.045007} {\bibfield  {journal}
  {\bibinfo  {journal} {Rev. Mod. Phys.}\ }\textbf {\bibinfo {volume} {94}},\
  \bibinfo {pages} {045007} (\bibinfo {year} {2022})}\BibitemShut {NoStop}%
\bibitem [{\citenamefont {Haapasalo}\ \emph {et~al.}(2011)\citenamefont
  {Haapasalo}, \citenamefont {Lahti},\ and\ \citenamefont
  {Schultz}}]{haapasalo11}%
  \BibitemOpen
  \bibfield  {author} {\bibinfo {author} {\bibfnamefont {E.}~\bibnamefont
  {Haapasalo}}, \bibinfo {author} {\bibfnamefont {P.}~\bibnamefont {Lahti}}, \
  and\ \bibinfo {author} {\bibfnamefont {J.}~\bibnamefont {Schultz}},\
  }\bibfield  {title} {\enquote {\bibinfo {title} {Weak versus approximate
  values in quantum state determination},}\ }\href
  {https://journals.aps.org/pra/abstract/10.1103/PhysRevA.84.052107} {\bibfield
   {journal} {\bibinfo  {journal} {Phys. Rev. A}\ }\textbf {\bibinfo {volume}
  {84}},\ \bibinfo {pages} {052107} (\bibinfo {year} {2011})}\BibitemShut
  {NoStop}%
\end{thebibliography}

%

\end{document}